\documentclass[12pt,aps,pre,preprint]{revtex4}

\usepackage{amssymb}
\usepackage{graphicx}

\begin{document}

\title{Evolution equation for tagged particle density and correlations in single file diffusion}
\author{Gonzalo Su\'arez, Miguel Hoyuelos and H\'ector O. M\'artin}

\affiliation{Departamento de F\'{\i}sica, Facultad de Ciencias Exactas y Naturales, Universidad Nacional de Mar del Plata and Instituto de Investigaciones F\'{\i}sicas de Mar del Plata (Consejo Nacional de Investigaciones Cient\'{\i}ficas y T\'{e}cnicas), Funes 3350, 7600 Mar del Plata, Argentina}

\begin{abstract}
We derive and study a theoretical description for single file diffusion, i.e., diffusion in a one dimensional lattice of particles with hard core interaction.  It is well known that for this system a tagged particle has anomalous diffusion for long times.  The novelty of the present approach is that it allows for the derivation of correlations between a tagged particle and other particles of the system at a given distance with empty sites in between.  The behavior of the correlation gives deeper insights into the processes involved.  Numerical integration of differential equations are in good agreement with Monte Carlo simulations.
\end{abstract}

\maketitle


\section{Introduction}

Single file diffusion, or dynamics, is the diffusion of $N$ identical particles in a one dimensional lattice, with the following features. Double occupancy is not allowed in each site of the lattice. In general, hard core interaction between particles is assumed. Particles can not jump one over the other, so that the order is preserved.
The most salient feature of this model is that, for large times, a tagged particle has anomalous diffusion, with mean square displacement (MSD) $\Delta^2 x \sim t^{1/2}$ \cite{harris}. This simple model, or different generalizations, has been useful to describe a wide variety of microscopic processes in physics, chemistry or biology (see, e. g., \cite{richards,karger,wei,chung}).

From a theoretical point of view, the model has motivated the development of many interesting results. A few examples can be mentioned.  Considering a lattice spacing $a$ and a homogeneous concentration per site $c$, Richards \cite{richards} demonstrated that, for large times, the asymptotic behavior of the MSD of a tagged particle is 
\begin{equation}
\Delta^2 x = \frac{2a^2(1-c)}{c} \left(\frac{\Gamma t}{\pi} \right)^{1/2},
\label{eq:1} 
\end{equation}
where $\Gamma$ is the transition rate of a particle to an empty neighboring site.
In \cite{beijeren}, van Beijeren \textit{et al} investigated the MSD for all times and for finite chains. R\"odenbeck \textit{et al} \cite{rodenbeck} studied the density defined in the space of all particle positions, with boundary conditions that take care of the hardcore interaction and, using the reflection principle, obtained an exact expression for the tagged particle density that slightly differs from a Gaussian distribution.  Kollmann \cite{kollmann} generalized previous results to interaction potentials, that exclude mutual passage, different from hardcore interaction. Lizana \textit{et al} \cite{lizana} obtained exact results for the density of the tagged particle in a finite system of length $L$ with reflecting boundaries for three different time regimes: normal diffusion for short times, anomalous diffusion for intermediate times, and constant MSD for times longer than the equilibrium time. Manzi \textit{et al} \cite{manzi} analyzed the role of different initial particle configurations of $k$-mers in the transition from normal to anomalous diffusion. Centres and Bustingorry \cite{centres} described the single-file diffusion using a discrete Edward-Wilkinson equation. Nelissen \textit{et al} \cite{nelissen} showed that the single file diffusion depends on the interaction between particles and damping mechanisms in the system. Lucena \textit{et al} \cite{lucena} showed that the diffusion in a quasi-one-dimensional system can be tunned by an attractive or repulsive potential.
 In order to have a better understanding of the processes that take place at atomic scale, there has been experiments carried out by Lutz \textit{et al} \cite{lutz1} and Delfau \textit{et al} \cite{delfau1}. The former studied the diffusion of particles  constrained in a circular optical trap; the latter analyzed the dynamics of particles confined in a linear channel of finite length. Both experiments witnessed the anomalous behavior for large times. More recently, Carvalho \textit{et al}\cite{carvalho} investigated the effect of a repulsive Yukawa potential in a quasi-one-dimensional system.

As far as we know, no diffusion equation for the density of a tagged particle has been proposed, probably because there is an important difficulty.  It is necessary to take into account the correlation between tagged and untagged particles, since the presence of the untagged particles modifies the diffusion making it anomalous. This correlation, in turn, is related to correlations among three particles, giving rise to an infinite hierarchy of equations.  An alternative to describe the tagged particle density with a diffusion equation is the fractional dynamics approach \cite{metzler}. However, the fractional diffusion equation can not represent the two different regimes, for short and long times, that are present in single file diffusion. Fractional Langevin equations can be used for single file diffusion, as analyzed in \cite{lizana2}, where finite size effects can be taken into account introducing a force and correlations between particles in the long time regime were also obtained.

The novelty of this paper is the derivation of a diffusion equation for the correlation $u_{ij}$ of having the tagged particle in site $i$ of the one-dimensional lattice and a particle (tagged or untagged) at distance $j$, with empty sites in between.  In the derivation it is necessary to make use of some appropriate approximations in order to avoid the infinite hierarchy.  We show that, despite the needed approximations, the equation correctly describes the two regimes of normal and anomalous diffusion for short and long times respectively.

\section{Diffusion equation and correlations}

In the following, we will focus on a one-dimensional infinite discrete lattice.  The lattice spacing is $a$, so that the position of a given site $i$ can be obtained by $x=a i$, $\forall i \in \mathbb{Z}$. The occupation number of each site is $\mathsf{m}_i$, equal to 1 or 0 depending on the site being occupied or empty respectively.  We define $\mathsf{n}_i$ as the occupation number corresponding to only a tagged particle, while $\mathsf{m}_i$ corresponds to any particle.  A particle jumps to a neighboring site with a transition rate $\Gamma$.  Due to a hard core interaction, the jump is performed only if the destination site is empty.  In the following, we use sans serif font for variables of a particular configuration of the system, and italic font for configuration averages, i.e., $n_i = \langle \mathsf{n}_i \rangle$ and $m_i = \langle \mathsf{m}_i \rangle$.

Let us consider the tagged particle current $\mathsf{J}_i$ between sites $i$ and $i+1$.  We can evaluate $\mathsf{J}_i$ by considering the particle configuration and the possible processes:
\begin{equation}
\mathsf{J}_i = \mathsf{n}_i (1 - \mathsf{m}_{i+1}) \Gamma - \mathsf{n}_{i+1}(1-\mathsf{m}_i) \Gamma.
\label{eq:j}
\end{equation}
The first term corresponds to a jump of the tagged particle from site $i$ to $i+1$, given that site $i+1$ is empty, and the second term represents a jump from $i+1$ to $i$. This equation can be rewritten as
\begin{equation}
\mathsf{J}_i = - (\mathsf{n}_{i+1} - \mathsf{n}_i) \Gamma + \mathsf{q}_i \Gamma,
\label{j2}
\end{equation}
where $\mathsf{q}_i = \mathsf{m}_i \mathsf{n}_{i+1} - \mathsf{n}_i \mathsf{m}_{i+1}$.  The densities at position $x$ are $m(x) = m_i/a$ and $n(x) = n_i/a$.  Similarly, we define $J(x) = \langle \mathsf{J}_i \rangle$ and $q(x) = \langle \mathsf{q}_i \rangle/a^2$.  From Eq. (\ref{j2}) we get
\begin{equation}
J = - D \frac{\partial n}{\partial x} + D q,
\label{j3}
\end{equation}
where $D=a^2 \Gamma$ is the diffusion coefficient, and we have approximated $\frac{n(x+a) - n(x)}{a} \simeq \frac{\partial n}{\partial x}$.  The diffusion equation is
\begin{equation}
\frac{\partial n}{\partial t} = -\frac{\partial J}{\partial x} = D\frac{\partial}{\partial x} \left(\frac{\partial n}{\partial x} - q  \right).
\label{diff}
\end{equation}
The anomalous diffusion at large times can only be produced by the correlations in $q$.  As a first approximation, we can decorrelate the products in $q$, i.e., 
\begin{equation}
\langle \mathsf{m}_i \mathsf{n}_{i+1} \rangle \simeq \langle \mathsf{m}_i \rangle \langle \mathsf{n}_{i+1} \rangle, \label{decorr}
\end{equation}
so we get
\begin{equation}
q \simeq a m \frac{\partial n}{\partial x} - a n \frac{\partial m}{\partial x}.
\label{eq:q}
\end{equation}
and
\begin{equation}
\frac{\partial n}{\partial t} = D\frac{\partial}{\partial x} \left( (1-am)\frac{\partial n}{\partial x} + a n \frac{\partial m}{\partial x}  \right) \ \ \ \ \mbox{(short times)}.
\label{diff2}
\end{equation}
For simplicity, we assume that the particles are uniformly distributed, so that $\frac{\partial m}{\partial x} = 0$.  The tagged particle density, $n$, has a normal diffusion with an effective diffusion coefficient $D' = D (1- a m)$.  This result holds for short times when starting from random initial distributions, i.e. while the decorrelation (\ref{decorr}) is a valid approximation.  

The crossover time between normal and anomalous behavior, $t_c$, can be obtained by comparing the MSD in both regimes.  For normal diffusion we have $\Delta^2 x = 2 D' t$, and for anomalous diffusion we have Eq. (\ref{eq:1}), with $c = am$, so 
\begin{equation}
t_c = \frac{1}{m^2 D \pi}
\label{tc}
\end{equation}

In the rest of the paper we focus on a discrete description of the system.  The discrete version of Eq. (\ref{diff}) is
\begin{equation}
\frac{\partial n_i}{\partial t} = \Gamma\left[ n_{i+1} + n_{i-1} - 2n_i - (q_i - q_{i-1})\right].
\label{diffd}
\end{equation}
For simplicity, we will consider a lattice spacing $a=1$.

In the following, we propose an approach that goes beyond the decorrelation of Eq. (\ref{decorr}) in order to obtain a description that holds in the anomalous regime.

\section{Tagged-untagged particle correlation}

It is convenient to define a new correlation that is directly related to $q_i=\langle \mathsf{q}_i \rangle$.
We define the tagged-untagged particle correlation, $u_{i,j}$, as the configuration average of the tagged particle density in site $i$ times the  particle density in site $i+j$ (tagged or untagged), with empty sites in between:
\begin{equation}
u_{i,j} = \langle \mathsf{n}_i (\mathsf{m}_{i+1}-1)... \mathsf{m}_{i+j} \rangle
\label{uij}
\end{equation}
with $|j| \geq 1$.  We assume average initial distributions of tagged and untagged particles that are symmetric respect to $i=0$.  In particular, in Sect. \ref{results} we discuss an initial random distribution of particles. We choose one of these particles as the tagged particle, and we define the position of particles, such that, $i=0$ corresponds to the initial position of this tagged particle. On average, the symmetry $i \rightarrow -i$ is preserved, so that $u_{i,j} = u_{-i,-j}$.  We make use of this symmetry in order to consider only positive values of $j$.  The relation with correlation $q_i$ is 
\begin{equation}
q_i = u_{i+1,-1} - u_{i,1} = u_{-i-1,1} - u_{i,1}.
\label{q2}
\end{equation}

The tagged particle density is related to $u_{i,j}$ by
\begin{equation}
n_i = \sum_{j=1}^\infty u_{i,j}.
\label{niuij}
\end{equation}
The interparticle distribution, $p_j$, gives the probability of finding a particle at distance $j$ from a tagged particle, with empty sites in between. It is given by 
\begin{equation}
p_j = \sum_{i=-\infty}^\infty u_{i,j}.
\end{equation}
Since the tagged particles are not different from the rest in their evolution, $p_j$ actually gives the interparticle distribution between any pair of consecutive particles.  It can be shown that the equation for $p_j$, for $j>1$ is
\begin{equation}
\frac{\partial p_j}{\partial t} = 2\Gamma \left[p_{j+1} + p_{j-1} - 2p_j + p_1 (p_j - p_{j-1}) \right],
\label{pj}
\end{equation}
and for $j=1$,
\begin{equation}
\frac{\partial p_1}{\partial t} = 2\Gamma\left[p_2 - p_1(1 - p_1) \right]
\end{equation}
We have assumed that the three body correlation function for two particles next to each other and a third one at distance $j$ can be replaced by the product $p_1 p_j$.
The stationary solution of these equations correspond to a homogeneous random distribution of particles with constant density $m_i = c$ and is given by $p_j^\mathrm{st} = c(1-c)^{j-1}$.

\begin{figure}%
\includegraphics[width=11cm]{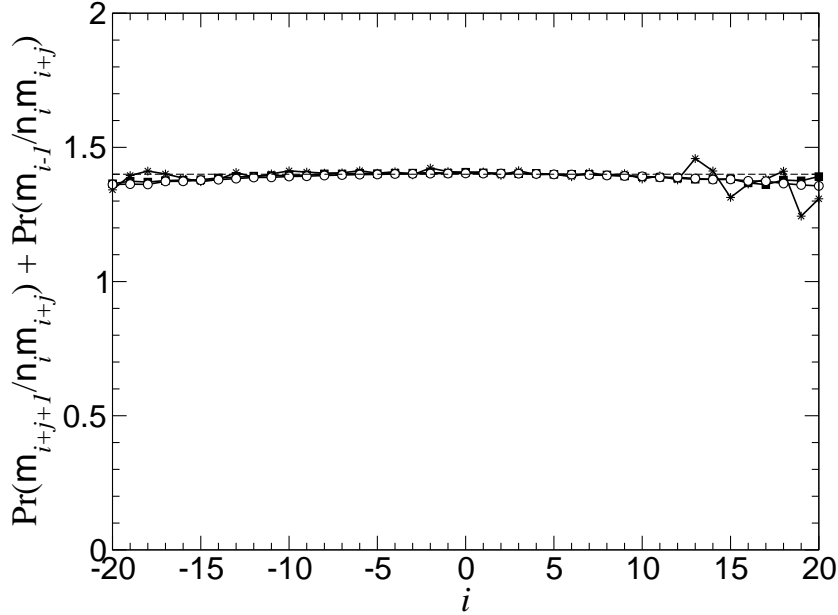}%
\caption{Numerical evaluation of the approximation of Eq. (\ref{aprsum}). The quantity plotted is $\mathrm{Pr}(\mathsf{m}_{i+j+1}/\mathsf{n}_i \mathsf{m}_{i+j}) + \mathrm{Pr}(\mathsf{m}_{i-1}/\mathsf{n}_i \mathsf{m}_{i+j})$ against position $i$ for different values of $j$: $j=1$ ($\circ$), $j=3$ ($\blacksquare$) and $j=6$ ($*$). A value of $c=0.7$ was used. The initial conditions are random, so that $2p_1 = 1.4$ for all times. The sum of the conditional probabilities, that individually depend on $i$ and $j$, coincides with the constant value $1.4$. The range of the horizontal axis is larger than the range in which the tagged particle density takes relevant values, between about $-10$ and $10$ (see Fig. \ref{nit}, $\Diamond$). Lattice length $L=1000$, time $t=10^4$, number of samples 450000.}%
\label{aprchk}%
\end{figure}

Now it is possible to derive a closed equation for $u_{i,j}$ that is useful to obtain $q_i$ and to get a common description for the normal and anomalous behaviors of single file diffusion.  In the derivation we have to consider all possible processes in a small time interval, $\delta t$, that may modify the value of $u_{i,j}$ at time $t$. Schematically (for $j>1$),
\begin{eqnarray}
u_{i,j}(t+\delta t) &=& u_{i,j} \left[1 - \mathrm{Pr}\left( {\bullet \atop \scriptstyle i}^{\displaystyle\curvearrowright} \cdots {\circ \atop  \scriptstyle i+j}  \right) -
\mathrm{Pr}\left( {\bullet \atop \scriptstyle i} \cdots {}^{{\displaystyle\curvearrowleft \atop \ }\atop \ } \!\! {\circ \atop  \scriptstyle i+j}  \right) \right. \nonumber \\
&& \left.
- \mathrm{Pr}\left( {}^{{\displaystyle\curvearrowleft \atop \ }\atop \ } \!\!{\bullet \atop \scriptstyle i} \cdots  {\circ \atop  \scriptstyle i+j}  \right)  - 
\mathrm{Pr}\left( {\bullet \atop \scriptstyle i} \cdots {\circ \atop  \scriptstyle i+j}\!\!^{\displaystyle\curvearrowright \atop \ }  \right)\right] \nonumber \\
&& + \mathrm{Pr}\left( {\bullet \atop \scriptstyle i-1}\!\!^{\displaystyle\curvearrowright \atop \ } \cdots {\circ \atop  \scriptstyle i+j}  \right) +
\mathrm{Pr}\left( {\bullet \atop \scriptstyle i} \cdots {}^{{\displaystyle\curvearrowleft \atop \ }\atop \ } \!\!\!\!\!\! {\circ \atop  \scriptstyle i+j+1}  \right) \nonumber \\
&& + \mathrm{Pr}\left( {}^{{\displaystyle\curvearrowleft \atop \ }\atop \ } \!\!\!{\bullet \atop \scriptstyle i+1} \cdots  {\circ \atop  \scriptstyle i+j}  \right) +
\mathrm{Pr}\left( {\bullet \atop \scriptstyle i} \cdots {\circ \atop  \scriptstyle i+j-1}\!\!\!\!\!^{\displaystyle\curvearrowright \atop \ }  \right),
\label{schem}
\end{eqnarray}
where `$\bullet$' represents a tagged particle, and `$\circ$' any kind of particle.  More explicitly, we have
\begin{eqnarray}
u_{i,j}(t+\delta t) &=& u_{i,j}\left[1 - 2\Gamma \delta t \right. \nonumber\\
&& \left.
- \Gamma \delta t (1 - \mathrm{Pr}(\mathsf{m}_{i-1}/\mathsf{n}_i \mathsf{m}_{i+j})) - \Gamma \delta t (1 - \mathrm{Pr}(\mathsf{m}_{i+j+1}/\mathsf{n}_i \mathsf{m}_{i+j})) \right] \nonumber \\
&& + u_{i-1,j+1} \Gamma \delta t + u_{i,j+1} \Gamma \delta t \label{expl} \\
&& + u_{i+1,j-1} \Gamma \delta t(1-\mathrm{Pr}(\mathsf{m}_i/\mathsf{n}_{i+1} \mathsf{m}_{i+j})) + u_{i,j-1} \Gamma \delta t(1-\mathrm{Pr}(\mathsf{m}_{i+j}/\mathsf{n}_i \mathsf{m}_{i+j-1})), \nonumber
\end{eqnarray}
where $\mathrm{Pr}(\mathsf{m}_{i-1}/\mathsf{n}_i \mathsf{m}_{i+j})$ is the conditional probability of having a particle in $i-1$ given that we have a tagged particle in $i$ and a particle in $i+j$, i.e.
\begin{equation}
\mathrm{Pr}(\mathsf{m}_{i-1}/\mathsf{n}_i \mathsf{m}_{i+j}) = \frac{\langle \mathsf{m}_{i-1} \mathsf{n}_i \mathsf{m}_{i+j} \rangle}{\langle \mathsf{n}_i \mathsf{m}_{i+j} \rangle}.
\label{cond}
\end{equation}
In order to get an approximation of this conditional probability, we observe that the presence of the untagged particle in $i+j$ has a weaker influence on this probability than the tagged particle in $i$.  We assume that this influence can be neglected, and approximate (\ref{cond}) by
\begin{equation}
\mathrm{Pr}(\mathsf{m}_{i-1}/\mathsf{n}_i \mathsf{m}_{i+j}) \simeq \mathrm{Pr}(\mathsf{m}_{i-1}/\mathsf{n}_i) = \frac{\langle \mathsf{m}_{i-1} \mathsf{n}_i \rangle}{\langle \mathsf{n}_i \rangle} = \frac{u_{-i,1}}{n_i}.
\label{condapr}
\end{equation}
In Eq. (\ref{expl}) we have also the conditional probability $\mathrm{Pr}(\mathsf{m}_{i+j+1}/\mathsf{n}_i \mathsf{m}_{i+j})$.  If we apply the same approximation, we obtain in this case a probability that is independent of the tagged particle position.  We propose a different approximation in order to keep the dependence on the tagged particle and also to obtain symmetric terms in the final equation for $u_{ij}$.  In this case we use
\begin{equation}
\mathrm{Pr}(\mathsf{m}_{i+j+1}/\mathsf{n}_i \mathsf{m}_{i+j}) + \mathrm{Pr}(\mathsf{m}_{i-1}/\mathsf{n}_i \mathsf{m}_{i+j}) \simeq 2 p_1,
\label{aprsum}
\end{equation}
so that
\begin{equation}
\mathrm{Pr}(\mathsf{m}_{i+j+1}/\mathsf{n}_i \mathsf{m}_{i+j}) \simeq 2 p_1 - \frac{u_{-i,1}}{n_i},
\label{condapr2}
\end{equation}
this last approximation turns out to be the simplest choice to get an equation for $u_{i,j}$ consistent with the equation for $p_j$ (\ref{pj}) when we sum over $i$  (see the appendix).  We numerically checked the validity of Eq. (\ref{aprsum}) and found an excellent agreement with the simulations, as shown in Fig. \ref{aprchk}.

Using (\ref{condapr}) and (\ref{condapr2}) in (\ref{expl}) we get the equation for $u_{i,j}$, for $j>1$,
\begin{eqnarray}
\frac{\partial u_{i,j}}{\partial t} &=& \Gamma \Biggl( u_{i-1,j+1} +  u_{i+1,j-1} + u_{i,j+1} + u_{i,j-1} - 4 u_{i,j}  \Biggr.  \nonumber \\
&&+ \Biggr. 2p_1 (u_{i,j} - u_{i,j-1}) + \frac{u_{-i,1} u_{i,j-1}}{n_i} - \frac{u_{-i-1,1} u_{i+1,j-1}}{n_{i+1}} \Biggr)
\label{uij2}
\end{eqnarray}
and the boundary condition for $j=1$,
\begin{equation}
\frac{\partial u_{i,1}}{\partial t} = \Gamma\left[ 2 u_{i,1}(p_1 - 1) + u_{i,2} + u_{i-1,2} \right].
\end{equation}
For random initial distributions of homogeneous concentration $c$ with one tagged particle in site $i=0$, the initial condition for $u_{i,j}$ is
\begin{equation}
\left. u_{i,j}\right|_{t=0} =  c(1-c)^{j-1} \delta_{i,0}.
\end{equation}

\section{Results}
\label{results}

In this section we present numerical simulation results of single file diffusion and compare them with the theoretical description given by Eq. (\ref{uij2}).

We consider a one dimensional lattice of $L$ sites with periodic boundary conditions. The lattice is large enough in order to avoid boundary effects during the time of the simulations.  The initial condition is a homogeneous random distribution of particles with density $c$, with one tagged particle in site $i=0$.  In this case, the interparticle distribution $p_j$ does not evolve in time, and is $p_j = p_j^\mathrm{st} = c(1-c)^{j-1}$ for all times (note that this is the same distribution as for the case of non-interacting particles). We are interested in the evolution of the tagged particle density, its MSD and the correlations. As shown below, the MSD was also evaluated for an ordered initial condition with equidistant particles.

\begin{figure}%
\includegraphics[width=11cm]{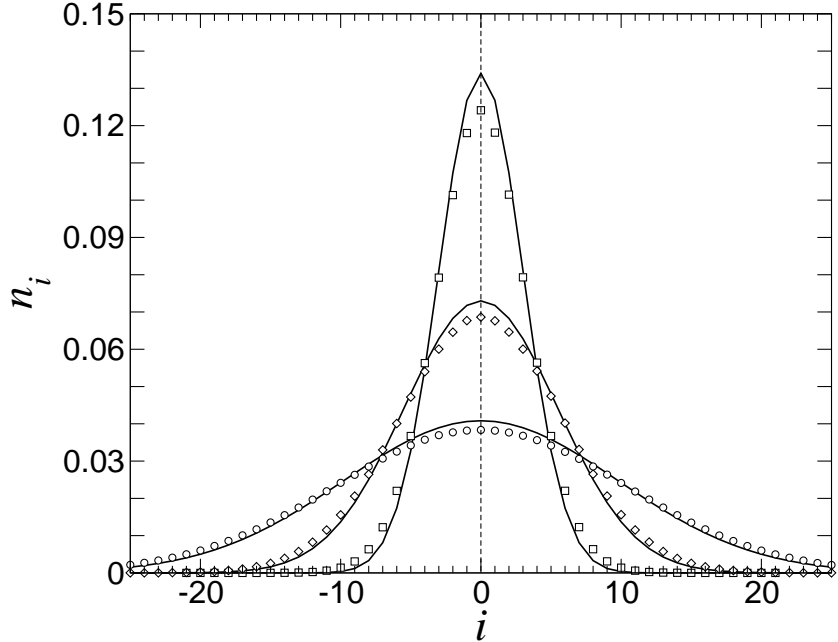}
\caption{\label{nit} Tagged particle density $n_i$, against position for different times in the anomalous regime: $t=10^3$ ($\Box$), $t=10^4$ ($\Diamond$) and $t=10^5$ ({\large$\circ$}). Concentration $c=0.7$ was used. Dots correspond to Monte Carlo simulations in a lattice of length $L=1000$; number of samples: 100000, and $\Gamma = 1/2$. The curves are the results of numerical integration of Eq. (\ref{uij2}) related to $n_i$ by (\ref{niuij}) \cite{numerical}. Each of the about 700 particles of each sample was considered as one tagged particle in order to reduce the statistical error.}
\end{figure}

Fig. \ref{nit} shows the tagged particle density, $n_i$, for different times in the anomalous regime.  We found a good agreement between Monte Carlo simulations (dots) and numerical integration of Eq. (\ref{uij2}) (curves).  The results can be approximately fitted with a Gaussian distribution whose variance increases with time as $t^{1/2}$.

\begin{figure}%
\includegraphics[width=11cm]{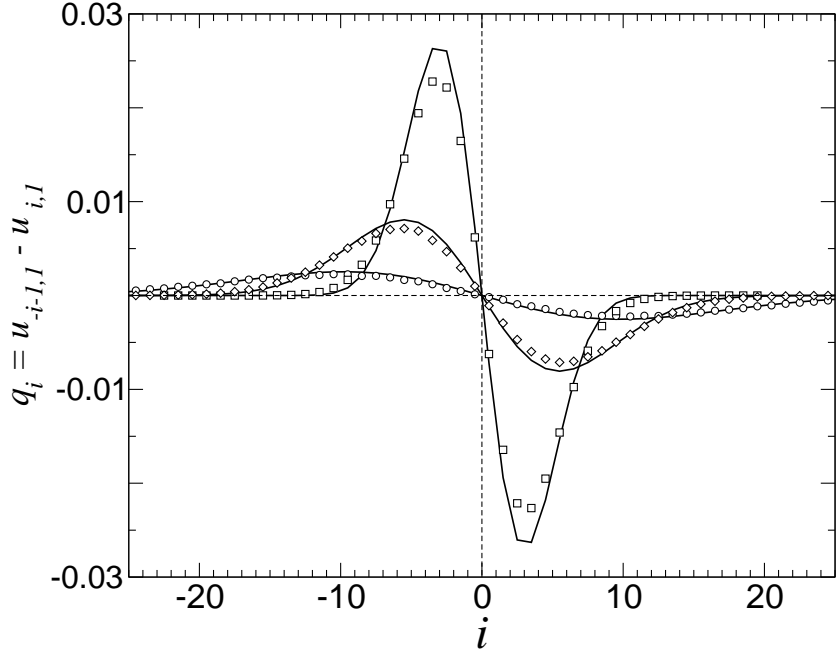}
\caption{Correlation $q_i=u_{-i-1,1} - u_{i,1}$ for different times. Same parameters as in Fig. \ref{nit}.}%
\label{qf}%
\end{figure}

In Fig. \ref{qf} we show the correlation $q_i$ for different times. Eq. (\ref{j3}) shows that this correlation is part of the particle current.  The physical meaning of $q_i$ is that of a current opposite to the free diffusion current, proportional to the concentration gradient.  The net effect is a reduction of the total current due to the interaction with the surrounding particles.  For short times (normal regime) this effect is manifested in a reduction of the effective diffusion coefficient: $D'= D(1-c)$.  For large times (anomalous regime) the effect of $q_i$ is more complex and it is not simply a reduction of the diffusion coefficient but a qualitative change in the temporal behavior of the MSD of $n_i$ that now increases as $t^{1/2}$.

\begin{figure}%
\includegraphics[width=11cm]{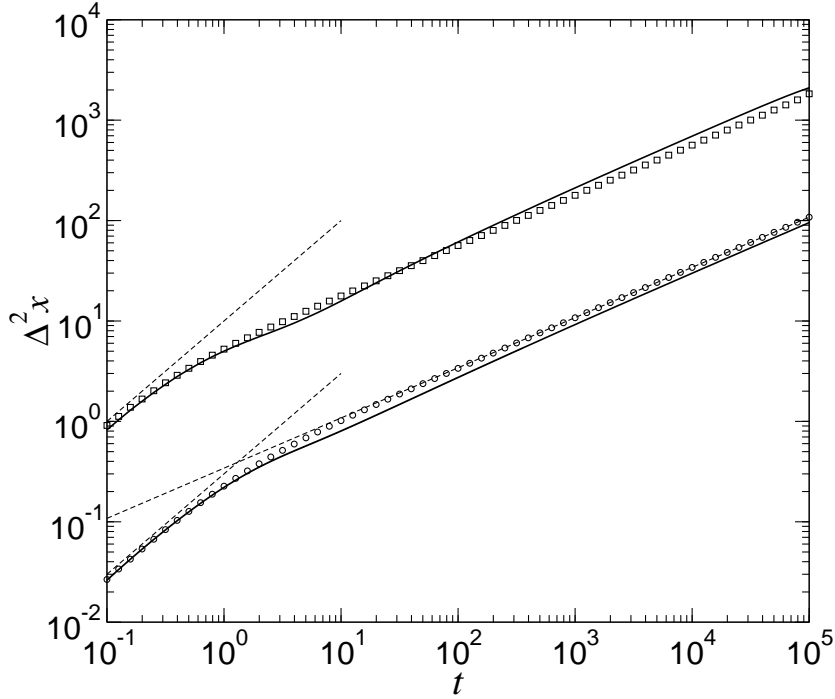}
\caption{Mean square displacement against time in log-log scale. Lower curve: random initial condition. Parameters as in Fig. \ref{nit}. Circles correspond to Monte Carlo simulations and the curve to numerical integration of Eq. (\ref{uij2}) related to $n_i$ by (\ref{niuij}). Dashed lines represent the asymptotic behaviors for short [$2\Gamma (1-c) t$] and long [Eq. (\ref{eq:1})] times, i.e., normal and anomalous regimes respectively. For times $10^{2}\le t\le 10^{5}$ the results obtained from numerical integration can be fitted by $\Delta^2 x = A t^{1/2}$, where the value of $A$ is $0.30$; the corresponding theoretical value from Eq. (\ref{eq:1}) is $0.34$.  Upper curve: initial condition with equidistant particles, concentration $c=0.5$ (lattice length 1000, number of samples 30000 and $\Gamma=1/2$).  The results were shifted one decade upwards for clarity. Squares correspond to Monte Carlo simulations and the curve to numerical integration of Eq. (\ref{uij2}). The dashed line represents the asymptotic behavior for short times, $2\Gamma t$, that holds for initially equidistant particles.}
\label{msd}%
\end{figure}

Fig. \ref{msd}, lower curve, shows the MSD against time in log-log scale for random initial conditions.  For short times we have the normal diffusion with $\Delta^2 x = 2\Gamma (1-c) t$. Results from numerical integration of Eq. (\ref{uij2}) agree with Monte Carlo simulations. The only effect of the approximations used in the previous section is that the MSD of the theoretical results is slightly smaller than the one of the Monte Carlo simulations. Other values of the concentration $c$ were considered and qualitatively the same results were obtained.  As the concentration is increased, the error introduced by the approximations of Eqs. (\ref{condapr}) and (\ref{aprsum}) also increses and the analytical results of the MSD in the anomalous regime are slightly more separated below the numerical results.  In the same figure, upper curve, we also show results for an initial condition with equidistant particles (the results are vertically shifted for clarity). Now, the behavior for short times is $\Delta^2 x = 2\Gamma t$, i.e, the tagged particle has normal diffusion in an effectively empty space, that is the reason why the factor $1-c$ is not present.  Anomalous diffusion appears for long times and the agreement between theory and simulations is similar to the case with random initial conditions.

\begin{figure}%
\includegraphics[width=11cm]{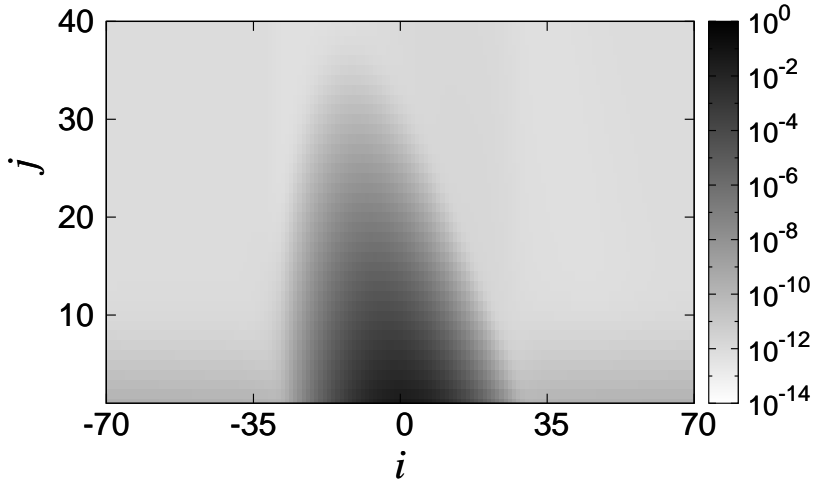}%
\caption{Integration results for the tagged-untagged particle correlation $u_{i,j}$ at $t=10^3$, for equidistant particles at $t=0$ with concentration $c=0.5$.  A logarithmic gray scale was used. The integration was performed in a lattice of size $-70\le i \le 70$ and $1 \le j \le 80$.  The method of integration is Runge-Kutta of order 5 with variable time interval.  A small initial value of order $10^{-9}$ was used for $i\neq 0$ in order to avoid numerical instabilities.}%
\label{fuij}%
\end{figure}

\begin{figure}%
\includegraphics[width=11cm]{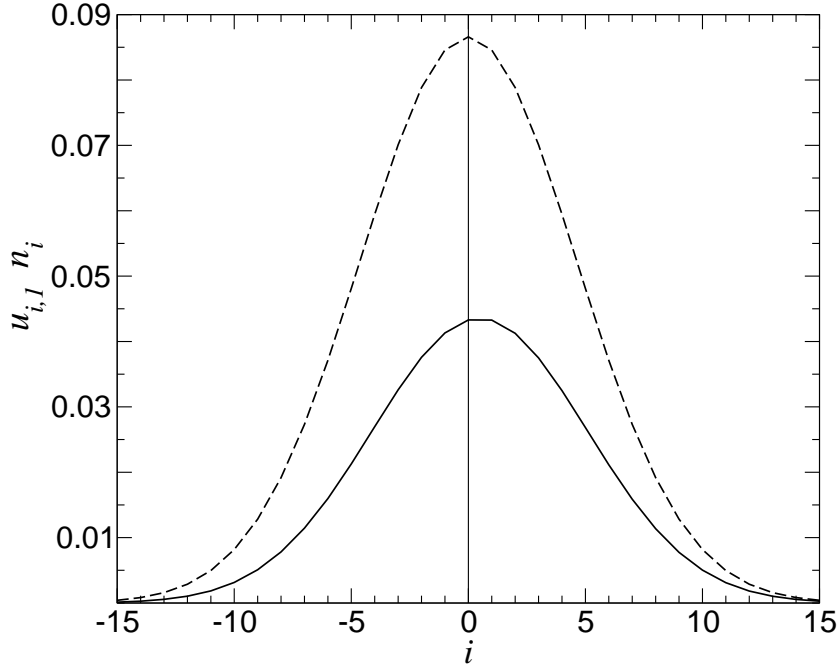}%
\caption{Integration results for $n_i$ (dashed curve) and $u_{i,1}$ (solid curve).  It can be seen that $u_{i,1}$ is slightly shifted to the right respect to $i=0$. Same parameters as in Fig. \ref{fuij}.}%
\label{ni_vs_ui1}%
\end{figure}

It is also interesting to analyze the asymmetric shape of $u_{i,j}$.  At $t=0$ the tagged particle is in $i=0$ and $u_{i,j}$ has a symmetric distribution with its maximum at $i=0$.  As time is increased, the distribution becomes tilted, and the maximum of $u_{i,j}$ is in the region $i>0$ for $j=1$ and in $i<0$ for large $j$. The reason is the following. For large $j$, it is more probable to find the tagged particle to the left of its original position ($i=0$) in order to give place to the $j-1$ empty spaces that are to the right.  On the other hand, for $j=1$, it is more probable to find the tagged particle in $i>0$ in order to have another particle adjacent to the right. The tagged-untagged particle correlation $u_{i,j}$ is shown in gray scale in Fig. \ref{fuij}. The figure shows integration results of Eq. (\ref{uij2}) at $t=10^3$ in the case of initial condition with equidistant particles (qualitatively the same results are found for other initial conditions and for large or intermediate times).  As explained before, the shape of $u_{i,j}$ is tilted to the left as $j$ is increased.  The case $j=1$ is more clearly seen in Fig. \ref{ni_vs_ui1}, were we show integration results for $u_{i,1}$ compared with $n_i$ for the same parameters as in Fig. \ref{fuij}; $u_{i,1}$ is shifted to the right respect to $i=0$, while $n_i$ is symmetric.  It is interesting to note that when summing over $j$, the tilts of $u_{i,j}$ to the right or to the left, for small or large $j$ respectively, cancel and the result is the symmetric density $n_i$.

\section{Conclusions}

We derived a theoretical description of single file diffusion that is different from previous approaches in one important aspect: it allows not only the derivation of the tagged particle density but also of the correlations between tagged and untagged particles.  We showed that the correlation given by $q_i=u_{-i-1,1} - u_{i,1}$ plays an important role in the diffusion equation for the tagged particle density, Eq. (\ref{diff}).  It is equivalent to a current, opposite to Fick's law current, that represents the impediment to diffusion generated by the surrounding particles. Knowing the behavior of correlations allows a deeper understanding of single file diffusion.

Despite some approximations needed to avoid an infinite hierarchy of equations, the equation obtained for the correlation $u_{i,j}$ correctly describes the normal and anomalous regimes for short and long times respectively.  This is particularly clear for the time dependence of the MSD of a tagged particle, for which the behaviors $\Delta^2 x \sim t$ (short times) and $\Delta^2 x \sim t^{1/2}$ (long times) were obtained.  Moreover, Eq. (\ref{uij2}) for $u_{i,j}$ holds for any initial condition with a symmetric tagged particle density, in particular we evaluated the MSD for equidistant particles and random initial conditions. 

Finally, let us note that, starting from a random distribution of particles, $p_1 = p_1^\mathrm{st} = c$ for all times.  As the number of collisions per unit time of a tagged particle is proportional to $p_1$, this number does not depend on time.  Then, the different behaviors of the MSD at short ($\Delta^2 x \sim t$) and long times ($\Delta^2 x \sim t^{1/2}$) are not due to the number of collisions as it was already pointed out in \cite{manzi}.  On the other hand, the combination of the number of collisions per unit time between a tagged particle at site $i$ and another particle at site $i+1$ ($\propto u_{i,1}$) and the number of collisions per unit time between a tagged particle at site $i+1$ and another particle at site $i$ ($\propto u_{i+1,-1}= u_{-i-1,1}$) plays a crucial role, because the difference $q_i = u_{-i-1,1} - u_{i,1}$ appears in the evolution equation of the tagged particle density $n_i$, Eq. (\ref{diffd}).  In other words, the frequency of collisions that take place at any site is irrelevant in the understanding of normal and anomalous behaviors, but a specific combination of frequency of collisions of particles in sites $i$ and $i+1$, when we take into account in which site the tagged particle is, plays a central role in the behavior of the MSD.

\section*{Appendix}

Here we explain in more detail the approximation of Eq. (\ref{condapr2}).  If we take the sum over $i$ in Eq. (\ref{expl}), we get
\begin{equation}
\sum_{i=-\infty}^\infty \frac{\partial u_{i,j}}{\partial t} = 2\Gamma \left[p_{j+1} + p_{j-1} - 2p_j + (a_j - a_{j-1})/2 \right],
\label{suma}
\end{equation}
where
\begin{equation}
a_j = \sum_{i=-\infty}^\infty u_{i,j} [ \mathrm{Pr}(\mathsf{m}_{i-1}/\mathsf{n}_i \mathsf{m}_{i+j}) + \mathrm{Pr}(\mathsf{m}_{i+j+1}/\mathsf{n}_i \mathsf{m}_{i+j}) ]
\label{aj0}
\end{equation}
Comparing (\ref{suma}) with Eq. (\ref{pj}) we obtain
\begin{equation}
a_j = 2p_1 p_j = \sum_{i=-\infty}^\infty u_{i,j} 2 p_1.
\label{aj}
\end{equation}
Now, using (\ref{aj0}) and (\ref{aj}), we find that the simplest choice to approximate the sum $\mathrm{Pr}(\mathsf{m}_{i-1}/\mathsf{n}_i \mathsf{m}_{i+j}) + \mathrm{Pr}(\mathsf{m}_{i+j+1}/\mathsf{n}_i \mathsf{m}_{i+j})$ is the one given in (\ref{aprsum}).

The approximations are also consistent with the equation for $n_i$ (\ref{diffd}) when we take the sum over $j$ of $u_{i,j}$.

\begin{acknowledgments}
This work was partially supported by Consejo Nacional de Investigaciones Cient\'{\i}ficas y T\'{e}cnicas (CONICET, Argentina, PIP 0041 2010-2012).  We would like to acknowledge to Prof. V. D. Pereyra for useful discussions.
\end{acknowledgments}


\begin{thebibliography}{00}
\bibitem{numerical} Numerical integration of Eq. (\ref{uij2}) was performed with a 4th order Runge Kutta method in a rectangular lattice of size 50 in the $i$ direction and 20 in the $j$ direction; time step: 0.01 for $t>100$ and 0.0001 for $t<100$.
\end{thebibliography}
\end{document}